\documentclass[twoside,slac_one,floatfix]{revtex4}
\usepackage{graphicx}
\usepackage{fancyhdr}
\usepackage{amsmath} 
\usepackage{bm}
\usepackage{amsxtra}
\usepackage{amssymb}
\usepackage{amsthm}
\usepackage{latexsym}
\usepackage{lscape}
\usepackage{float}

\begin{document}

\title{A search for charged massive long-lived particles at D0}

%

\author{J. Alimena for the D0 Collaboration}
\affiliation{Department of Physics and Astronomy, Brown University, Providence, RI, USA}

\begin{abstract}
We report on a search for charged massive long-lived particles (CMLLPs), based on 5.2 fb$^{-1}$ of data collected with the D0 detector at the Fermilab Tevatron $p\bar{p}$ collider. CMLLPs are predicted in many theories of physics beyond the standard model. We look for events in which one or more particles are reconstructed as muons but have speed and ionization energy loss $dE/dx$ inconsistent with muons produced in beam collisions. We present 95\% C.L. upper limits on the production cross section for long-lived scalar taus, for long-lived charginos in two SUSY scenarios, and for long-lived scalar top quarks. We also present exclusion mass ranges for the chargino and scalar top quark models.
\end{abstract}

\maketitle

\thispagestyle{fancy}


\section{Introduction}\label{sec:intro}
We report on a new search for massive particles, which are electrically
charged and have a lifetime long enough to escape the D0 detector
before decaying. Charged massive long-lived particles (CMLLPs) are
not present in the standard model (SM) nor are their distinguishing
characteristics (slow speed, high $dE/dx$) relevant for most physics
studies. Although the distinctive signature in itself provides sufficient motivation for
a search, some recent extensions to the SM suggest that CMLLPs exist
and are not yet excluded by cosmological limits~\cite{cosmo1,cosmo2}.
Indeed, our present model of big bang nucleosynthesis (BBN) has difficulties
in explaining the observed lithium production.
The existence of a CMLLP that decays during or after the time of BBN could resolve this disagreement~\cite{BBN}. 

We derive cross section limits for CMLLPs and compare them to theories
of physics beyond the SM. Supersymmetric (SUSY) models can predict either
the lightest chargino or the lightest scalar tau lepton (stau) to
be a CMLLP. We study a gauge-mediated supersymmetry breaking (GMSB)
model in which the next-to-lightest supersymmetric particle (NLSP)
is a long-lived stau lepton~\cite{NLSP2,NLSP3,stau-NLSP}. Other
explored models predict a light chargino whose lifetime
can be long if its mass differs from the mass of the lightest neutralino
by less than about 150 MeV~\cite{AMSB1,AMSB2}. This can occur
in models with anomaly-mediated supersymmetry breaking (AMSB) or in
models that do not have gaugino mass unification. There are two general
cases, where the chargino is mostly a higgsino and where the chargino
is mostly a gaugino, which we treat separately.

There are some SUSY models that predict a scalar top quark (stop)
NLSP and a gravitino LSP. These stop quarks hadronize into both charged
and neutral mesons and baryons that live long enough to be CMLLP
candidates~\cite{GMSB3}. Further, hidden valley models predict GMSB-like
scenarios where the stop quark acts like the LSP and does not decay,
but hadronizes into charged and neutral hadrons that escape the detector~\cite{HiddenValley1,HiddenValley2}.
In general, any SUSY scenario where the stop quark is the lightest
colored particle (which will happen in models without mass unification
and heavy gluinos) can have a stop CMLLP. Any colored CMLLP will have
additional complications of hadronization and charge exchange during
nuclear interactions, which we discuss below.

This search utilizes data collected between 2006 and 2010 with the
D0 detector~\cite{d0det} at Fermilab's 1.96 TeV $p\bar{p}$ Tevatron
Collider, and is based on 5.2 fb$^{-1}$ of integrated luminosity.
We reported earlier~\cite{D0 MSP PRL} on a similar 1.1 fb$^{-1}$
study, searching for events with a pair of slow-moving massive charged
particles. In addition to using the larger data sample, the present
search looks for one or more CMLLPs, characterized by high $dE/dx$
as well as by slow speed. Further, we explore predictions for long-lived
stop quarks as well as for stau leptons and charginos, which were
considered earlier. Other searches for long-lived particles include
those from the CDF Collaboration~\cite{CDF Run1 PRL,CDF Run2 PRL},
the CERN $e^{+}e^{-}$$\,$Collider LEP~\cite{LEP searches},
and the CERN $pp\,$ collider LHC~\cite{CMS HSCP paper,ATLAS cmllp paper}.

\section{The D0 Detector}\label{sec:detector}

The D0 detector~\cite{d0det} includes an inner tracker with two
components: an innermost silicon microstrip tracker (SMT) and a scintillating
fiber detector. We find the $dE/dx$ of a particle from the energy losses
associated with its track in the SMT. The tracker is embedded within
a 1.9 T superconducting solenoidal magnet. Outside the solenoid is a
uranium/liquid-argon calorimeter surrounded by a muon spectrometer,
consisting of drift tube planes on either side of a 1.8 T iron toroid.
There are three layers of muon detector planes: the A-layer, located
between the calorimeter and the toroid, and the B- and C- layers,
outside the toroid. Each layer includes planes of scintillation counters
which serve to veto cosmic rays. Thus the muon system provides multiple
time measurements from which the speed of a particle may be calculated.

\section{Important Variables}\label{sec:variables}

\begin{figure*}[tbp]
\noindent \begin{centering}
\includegraphics[scale=0.4]{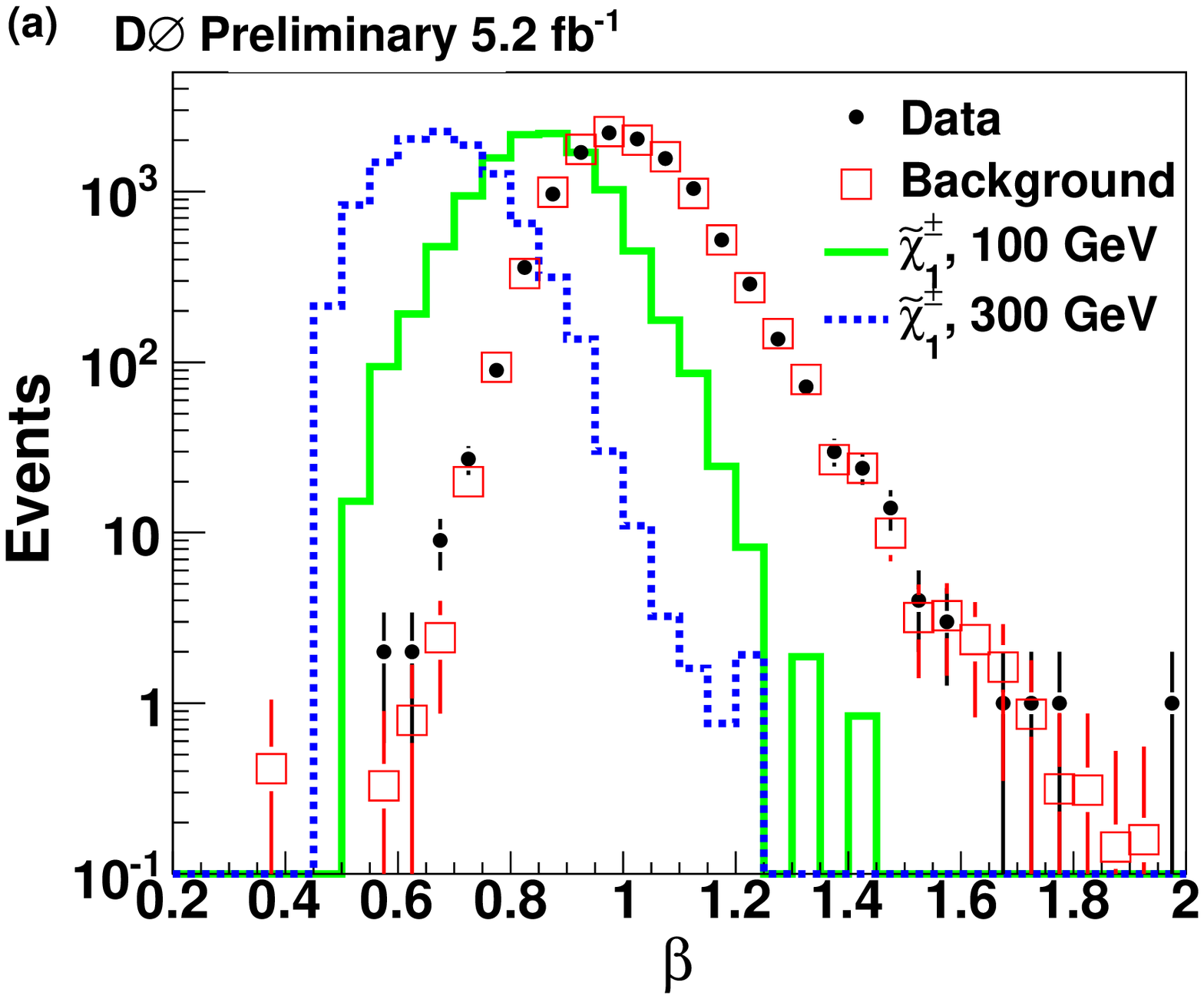}
\includegraphics[scale=0.4]{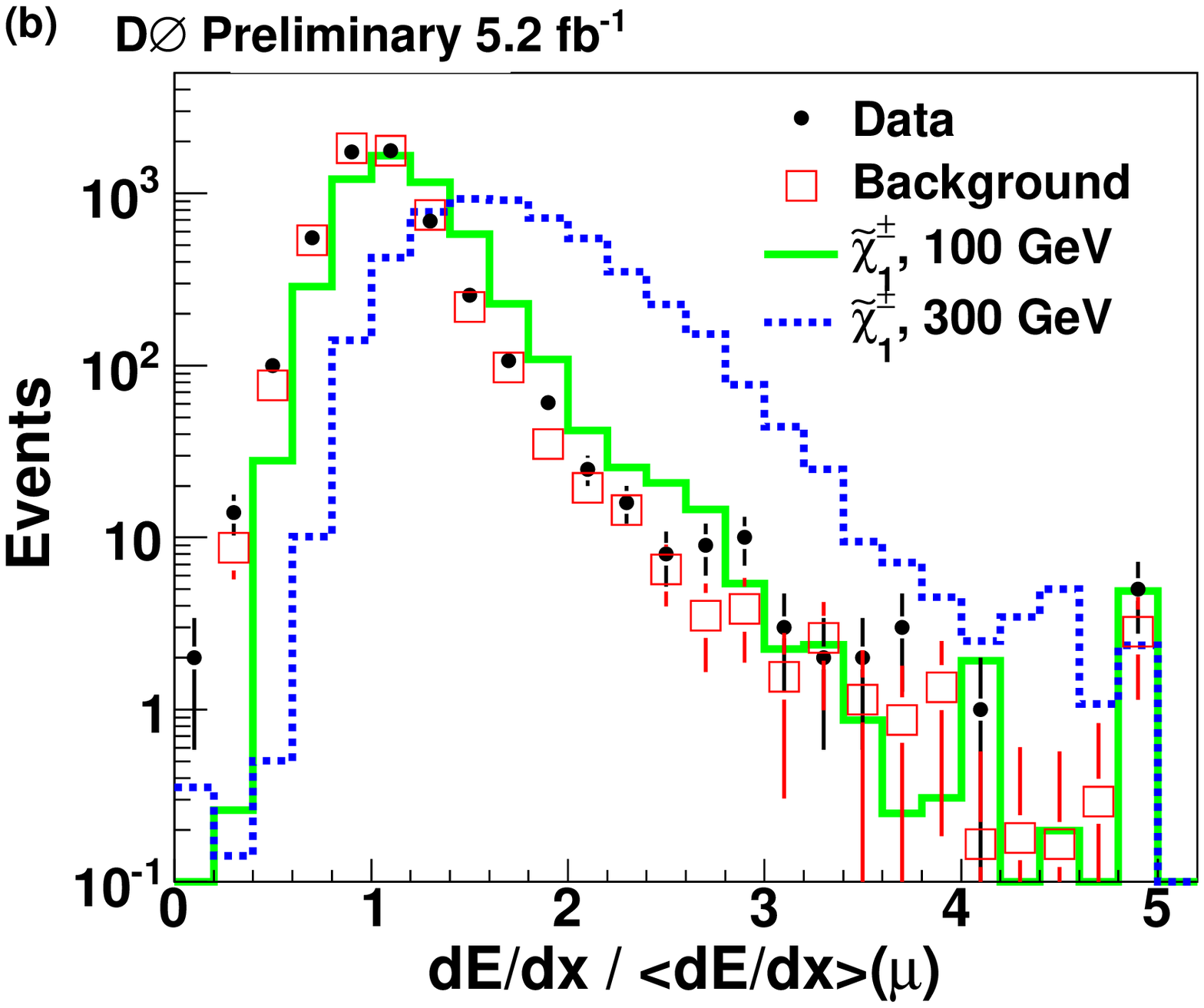}
\par\end{centering}
\caption{\label{fig:beta_dedx}Distributions of (a) speed ($\beta$) and (b)
$dE/dx$ for data, background, and signal (gaugino-like charginos
with a mass of 100 and 300 GeV) that pass the selection criteria.
The histograms have been normalized to have the same numbers of events.
We have adjusted the scale of the $dE/dx$ measurements so that the
$dE/dx$ of muons from $Z\rightarrow\mu\mu$ events peak at 1. All
entries exceeding the range of the histogram are added to the last
bin.}
\end{figure*}

Because we distinguish CMLLPs solely by their speed ($\beta$) and
$dE/dx$, we must measure these values in each event as accurately
as possible. Muons from $Z\rightarrow\mu\mu$ events studied throughout
the data sample allow us to calibrate the time measurement to better
than 1 ns, with resolutions between 2-4 ns, and to maintain the mean
$dE/dx$ constant to within 2\% over the data-taking period. From
a specific muon scintillation counter we calculate a particle's speed
from the time recorded and the counter's distance from the production
point, and we compute an overall speed from the weighted average of
these individual speeds, using measured resolutions. The ionization
loss data from the typically 8-10 individual hits in the SMT are combined
using an algorithm that omits the largest deposit to reduce the effect
of the Landau tail and corrects for track crossing angle. We calibrate
the $dE/dx$ measurements so that the $dE/dx$ of muons from $Z\rightarrow\mu\mu$
events peak at 1. Figure~\ref{fig:beta_dedx} shows the distributions
in $\beta$ and $dE/dx$ for data, background, and signal (defined in Section~\ref{sec:samples}).

\section{Event Selection}\label{sec:selection}
Selection of a candidate CMLLP event begins immediately at the time
of the interaction. Because of the high collision rate, we employ
a three-level trigger system to reduce the event rate to the 200 Hz
that can be recorded. The trigger system bases its decisions on characteristics
of the event, which for the CMLLP candidates is the presence of a
muon with a high momentum transverse to the beam direction $(p_{T})$.
In order to reduce triggers on cosmic rays, there is a time window
at the initial trigger level. This trigger gate reduces the trigger
efficiency by 10\% for CMLLPs with a mass of 300 GeV (as they will
be slow and some will be out-of-time) and contributes significantly
to our overall acceptance. We avoid a tighter timing gate usually
imposed at the second level of the muon trigger by accepting an alternative
requirement that the muon have a matching track in the SMT.

In the standard event reconstruction CMLLPs would appear as muons.
Thus, we select events with at least one well
identified high $p_{T}$ muon. For a useful $\beta$ measurement,
the event must have scintillator hits in the A-layer and either
the B- or C-layer, and for a valid $dE/dx$ measurement, we require at least
three hits in the SMT. For an optimal tracking and momentum measurement
we require the muon to be central, i.e., with a pseudorapidity\footnote{
The D0 coordinate system is cylindrical with
the $z$-axis along the proton beam direction, and the polar and azimuthal
angles are denoted by $\theta$ and $\phi$, respectively. The pseudorapidity
is defined as $\eta=$-ln{[}tan$\left(\theta/2\right)${]}.}
$ 
$$\mid\eta\mid<1.6$$\,$. To reject muons from meson decays we
impose the isolation requirement that the sum of the $p_{T}$ be less
than 2.5 GeV for all other tracks in a cone of radius$\:${\cal R}
$=\sqrt{(\Delta\phi)^{2}+(\Delta\eta)^{2}}<0.5$. We also require
that the total transverse calorimeter energy in an annulus of radius
$\:$$0.1<\:${\cal R}$\:<0.4$ about the muon direction be less
than 2.5 GeV. A requirement that the coordinate along the beam direction
of the distance of closest approach of the muon track to the beam
axis be $<40$ cm ensures that the particle passes through the SMT.

We impose criteria to eliminate cosmic rays. To select muons traveling
outwards from the apparent interaction point, we require that its
C-layer time be significantly greater than its A-layer time. We require
also that the muon's distance-of-closest-approach to the beam line
be less than 0.02 cm. These criteria are also applied to a second
muon in the event, if a second muon is present. In addition, for events with two muons we require
that the absolute value of the difference between each muon's A-layer
times be less than 10 ns. To reject cosmics that appear as two back-to-back
muons, we require for their pseudo-acolinearity $\Delta\alpha=|\Delta\theta+\Delta\phi-2\pi|>0.05$.

Events with a muon from a $W$ boson decay, with mismeasurements providing
inaccurate values of the muon's $\beta$ and $dE/dx$, constitute
a potentially large background. To study selection criteria for CMLLPs,
we calculate the transverse mass\footnote{The transverse mass is defined by $M_{T}=\sqrt{(E_{T}+{\not}E_{T})^{2}-(p_{x}+{\not}E_{x})^{2}-(p_{y}+{\not}E_{y})^{2}}$ 
where $E_{T}$ is the total energy transverse to the axis of the colliding
beams and ${\not}E_{T}$ is the total unbalanced or missing transverse
energy.} $M_{\text{T}}$,
and select data with $M_{\text{T}}<200$ GeV to model the data in
the absence of signal\footnote{The requirement $M_{T}<200$ GeV is customarily
used to select $W$ events.}. 
We choose selection
criteria that minimize the number of events surviving from this background
sample compared to events from simulations of CMLLP signal. We require
that events contain at least one muon with $p_{T}>60$ GeV. 
From a separate sample of muons from
$Z\rightarrow\mu\mu$ decays, we see that the association of a spurious scintillator hit at times gives an anomalously slow $\beta$ value. 
We use an algorithm that discards such hits through
minimizing the $\chi^{2}/d.o.f.$ for the $\beta$ calculated from
the different scintillator layers. By comparing the effect on the
background sample with the effect on simulated signal, we choose to
eliminate events unless the minimized speed $\chi^{2}/d.o.f.<2$. Finally,
we compare the track direction of the muon measured in the muon system
with that measured in the central tracker, and eliminate several events
with clearly mismatched tracks.

\section{Data and Monte Carlo Samples}\label{sec:samples}

To simulate potential signal events, we generate CMLLP candidates
using {\sc pythia}~\cite{Pythia}. Samples are generated with {\sc pythia}
for the long-lived stau lepton and chargino models, and the long-lived
stop quarks are hadronized using specialized routines interfaced with {\sc pythia}\footnote{The code for hadronizing stop quarks is at http://projects.hepforge.org/pythia6/examples/main78.f}.
Because the signature of the CMLLP cascade decays is model dependent
and difficult to simulate accurately, we generate direct pair-production
of the CMLLPs, without including cascade decays. We use the full D0
detector {\sc geant}~\cite{geant} simulation, which includes overlaid multiple
$p\bar{p}$$ $ interactions, to determine the
detector response for these samples. The results are applicable to models
with pair-produced CMLLPs with similar kinematics.

The stop quarks are distinct since they appear in charged or uncharged stop hadrons,
which may flip their charge as they pass through the detector. In the
simulation approximately 60\% of stop hadrons are charged following
initial hadronization~\cite{stop-fairbairn,stop-makeprang}, i.e.,
84\% of the events will have at least one charged stop hadron. Further,
stop hadrons may flip their charge through nuclear interactions as
they pass through material. We assume that stop hadrons have a probability
of 2/3 of being charged after multiple interactions and that anti-stop
hadrons have a probability of 1/2 of being charged~\cite{stop-fairbairn,stop-makeprang}.
For this analysis we require the stop hadron to be charged before and
after passing through the calorimeter, i.e., to be detected both in
the tracker and in the A-layer, and to be charged after the toroid,
i.e., to be detected in the B- or C-layers. The probability for at
least one of the stop hadrons to be detected is then 38\%, or 84\%
if charge flipping does not occur. We include these numbers as normalization
factors in the confidence level analysis discussed below.

Our final selection criteria is that the candidate's speed $\beta<1.$
Thus, we describe the background by the $\beta<1$ data events with
$M_{T}<200$ GeV, and search for CMLLP candidates in $\beta<1$ data
with $M_{T}>200$ GeV. We normalize the background and data samples
in the $\beta>1$ signal-free region. Because the uncertainties in
the speed measurements depend on the particle's $\eta$, due to detector
geometry, and the distributions in $\eta$ of the muons in the $M_{T}<200$
GeV sample differs from those in the $M_{T}>200$ GeV sample, we use
the signal-free region to derive a reweighting of the background sample
that matches its $\eta$ distribution to that of the data.

\section{Final Discriminants}\label{sec:discriminants}

\begin{figure*}[tbp]
\noindent \begin{centering}
\includegraphics[scale=0.4]{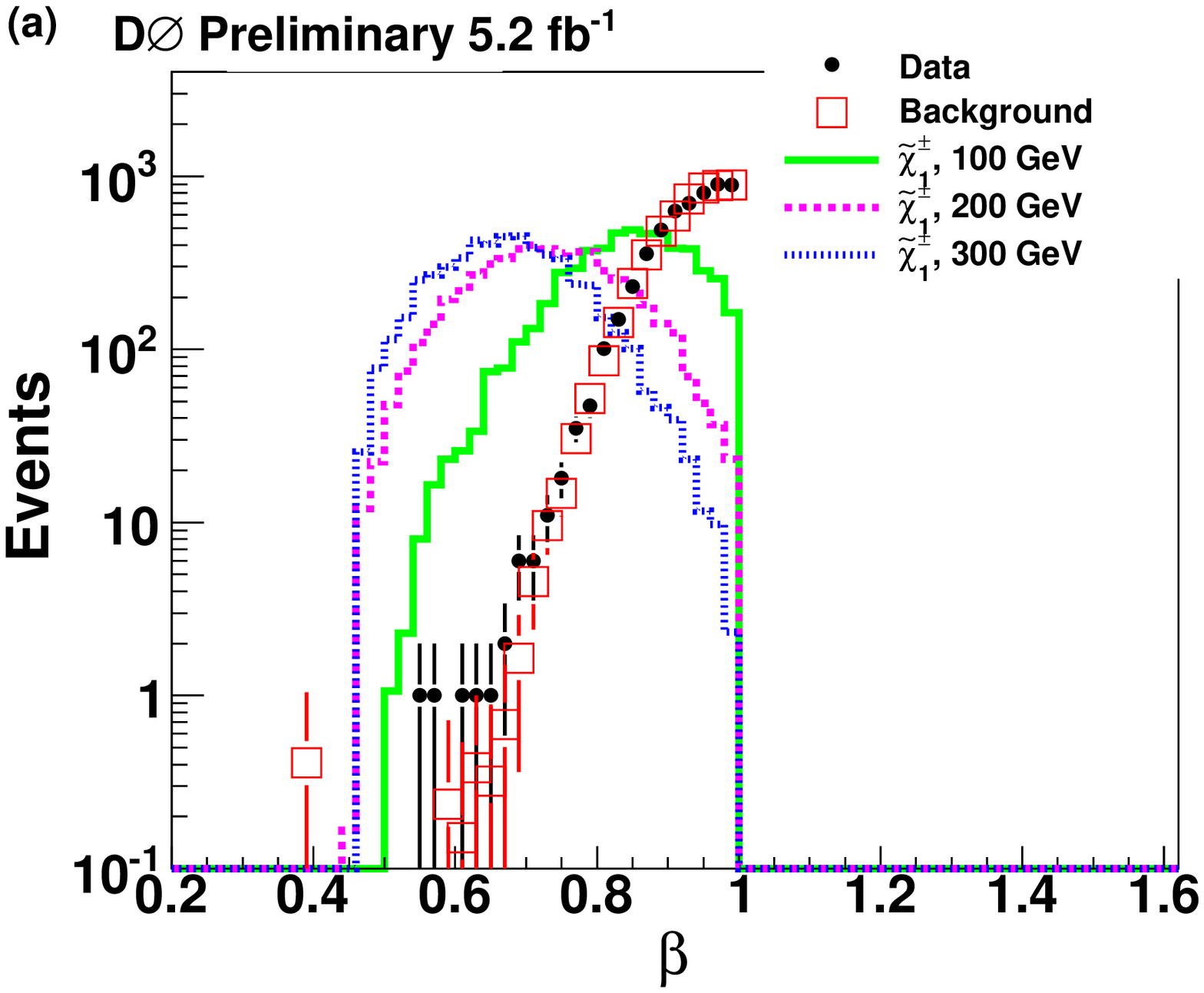}
\includegraphics[scale=0.4]{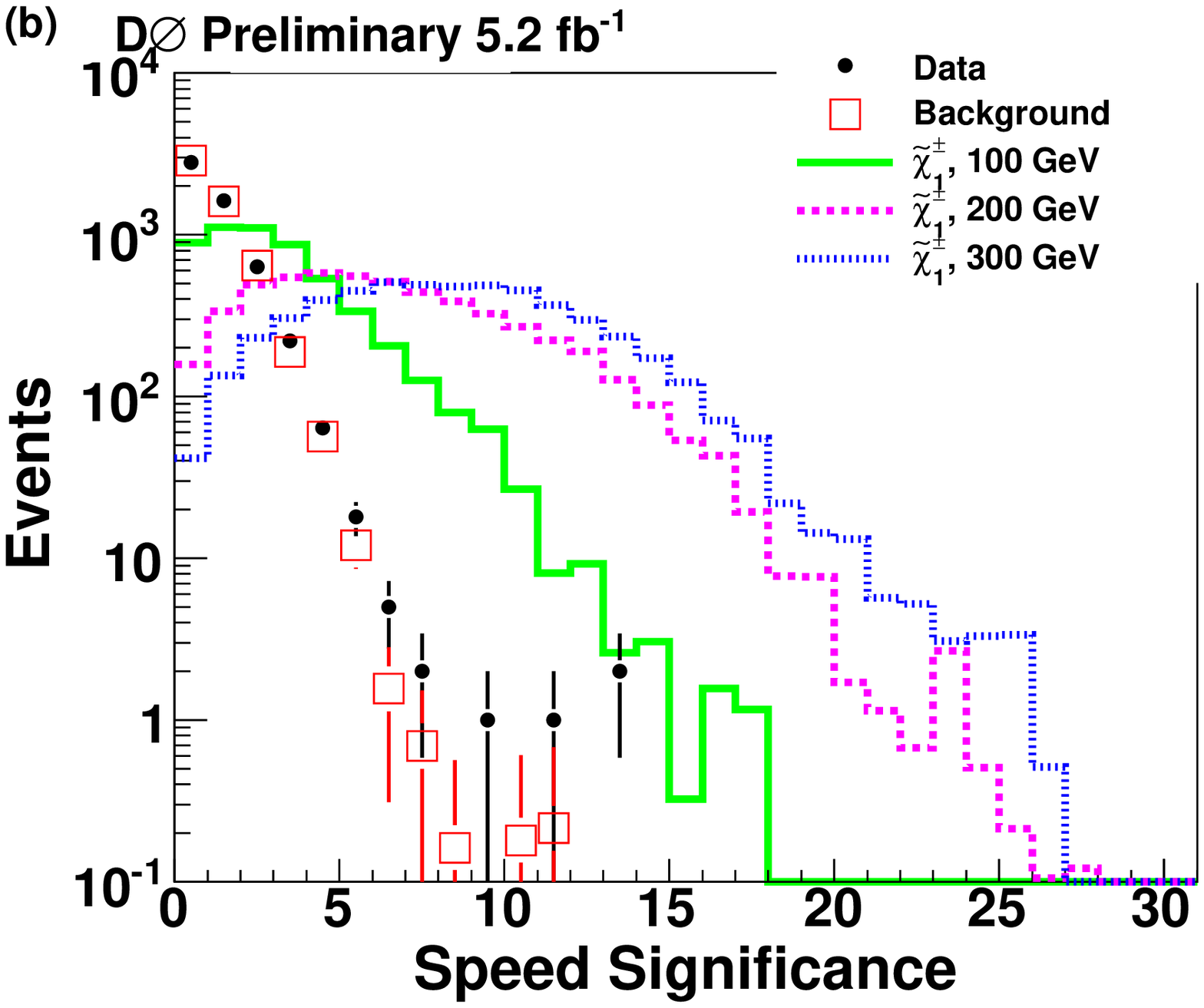}
\includegraphics[scale=0.4]{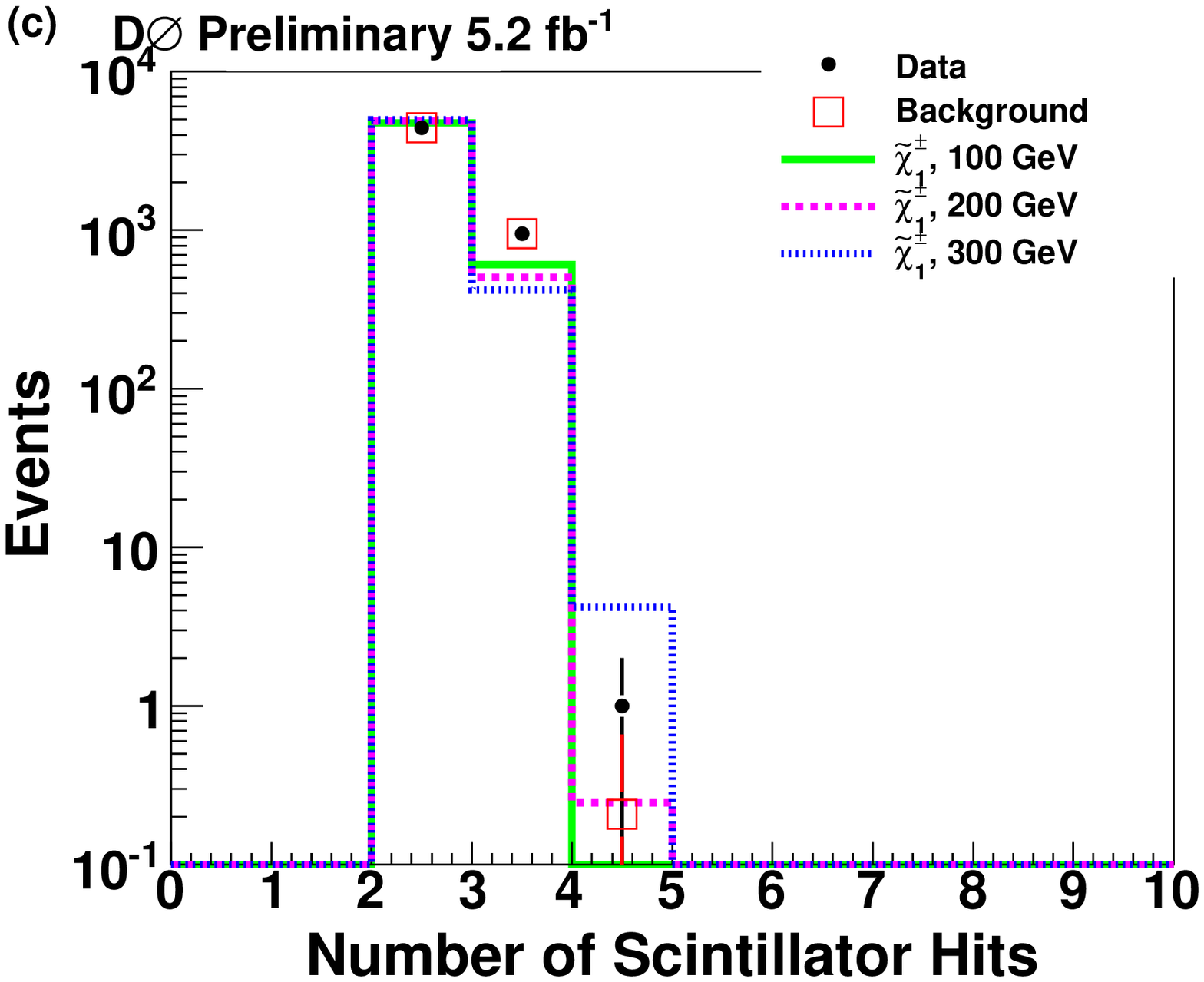}
\par\end{centering}
\caption{\label{fig:speed_finaldist} Final distributions related to the speed
for signal (100, 200, and 300 GeV gaugino-like charginos), background,
and data. The speed distribution (a), speed significance distribution
(b), and number of scintillator hits distribution (c). For
each plot, the histograms have been normalized to have the same number
of events.}
\end{figure*}

\begin{figure*}[tbp]
\noindent \begin{centering}
\includegraphics[scale=0.4]{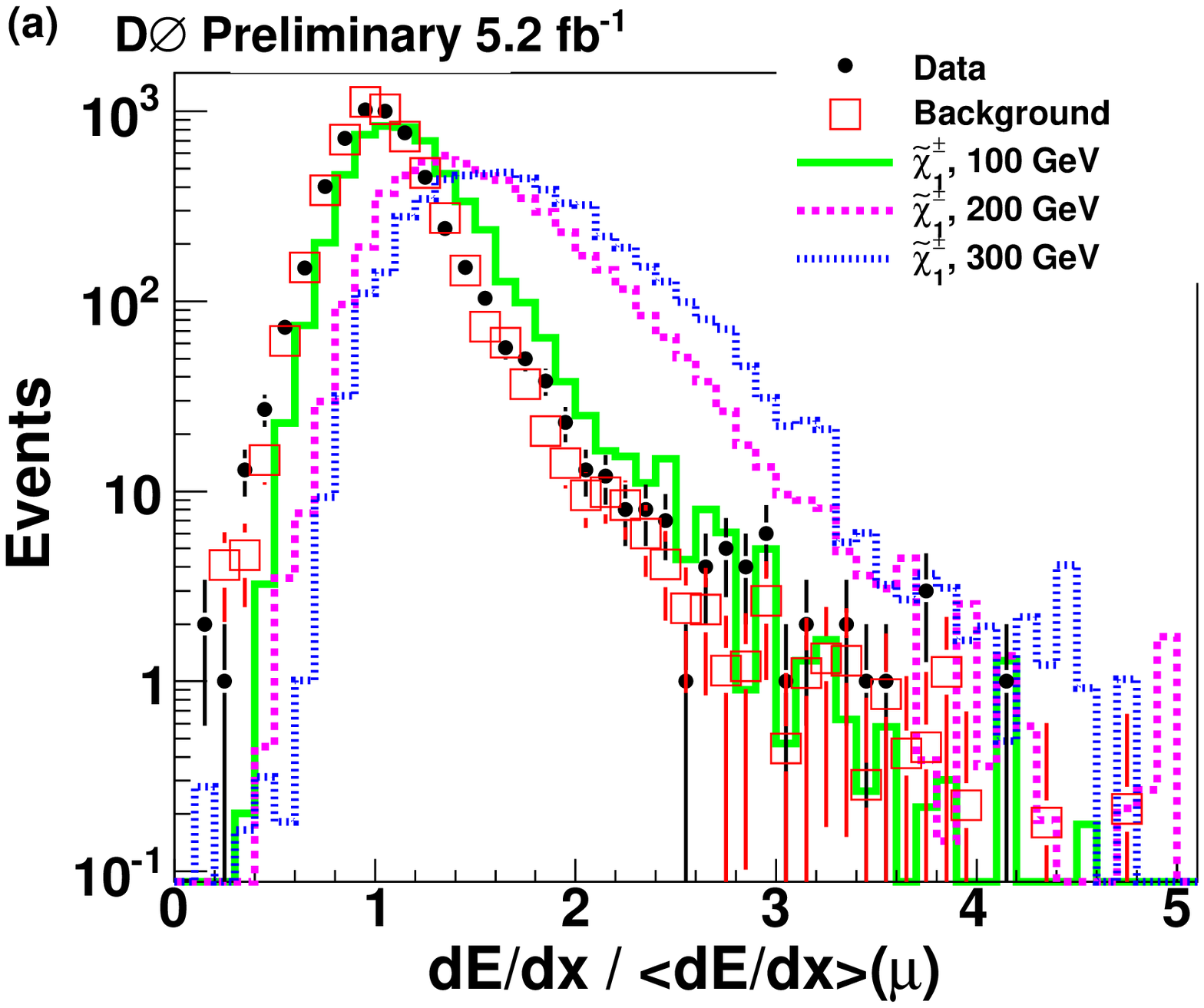}
\includegraphics[scale=0.4]{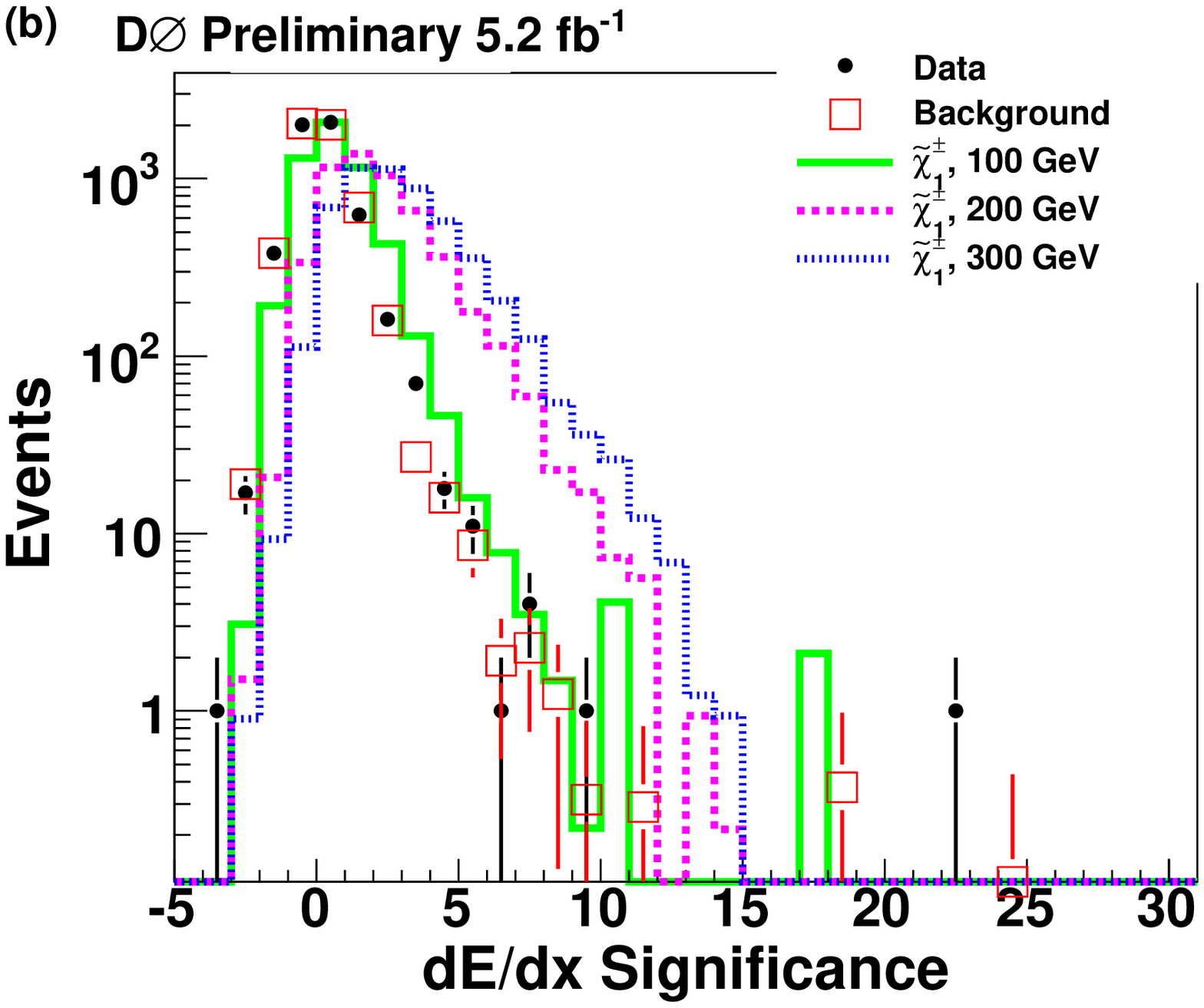}
\includegraphics[scale=0.4]{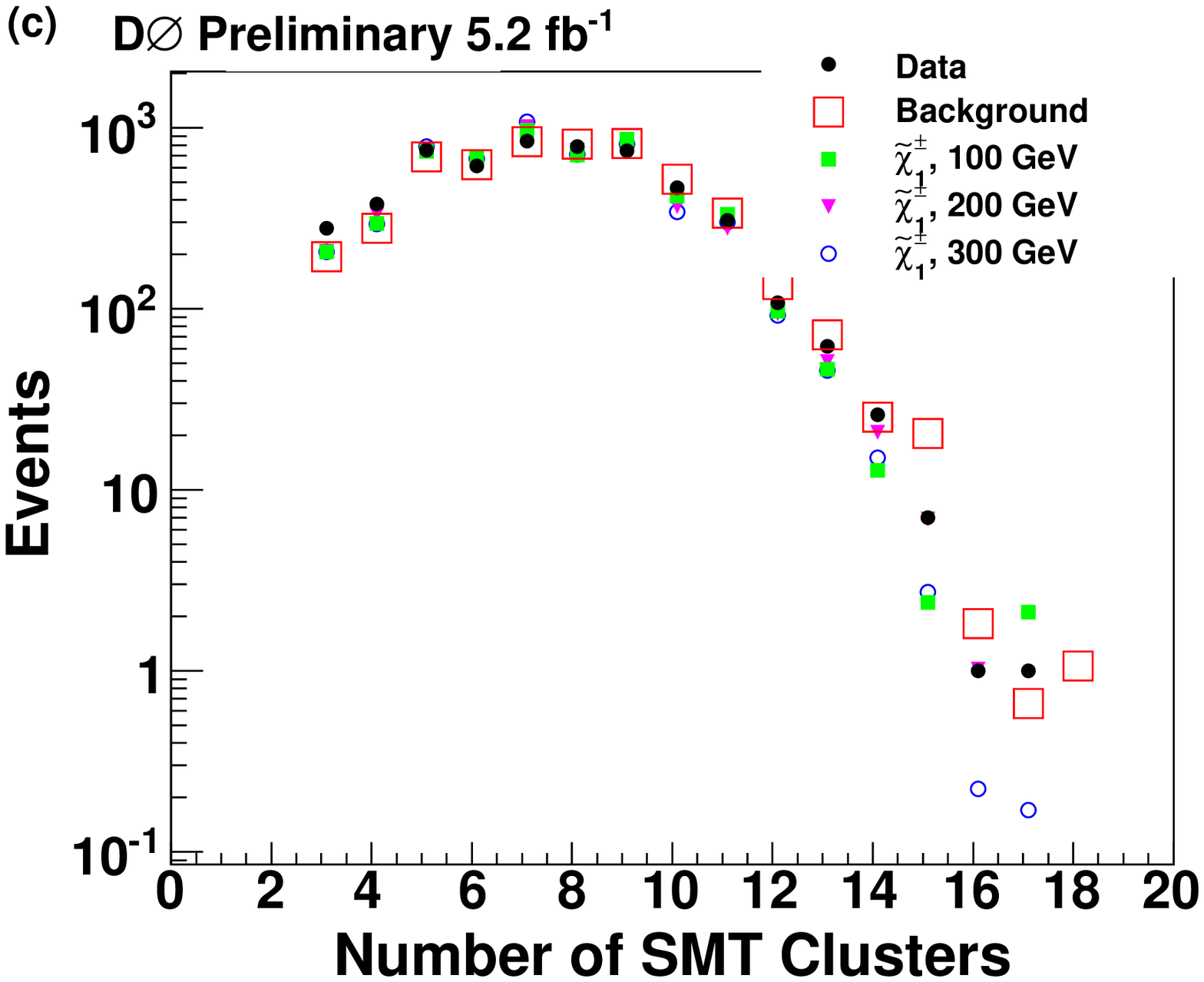}
\par\end{centering}
\caption{\label{fig:dedx_finaldist} Final distributions related to the $dE/dx$
for signal (100, 200, and 300 GeV gaugino-like charginos), background,
and data. The $dE/dx$ distribution (a), $dE/dx$ significance
distribution (b), and number of SMT clusters distribution (c).
For each plot, the histograms have been normalized to have the same
number of events.}
\end{figure*}

\begin{figure*}[tbp]
\noindent \begin{centering}
\includegraphics[scale=0.4]{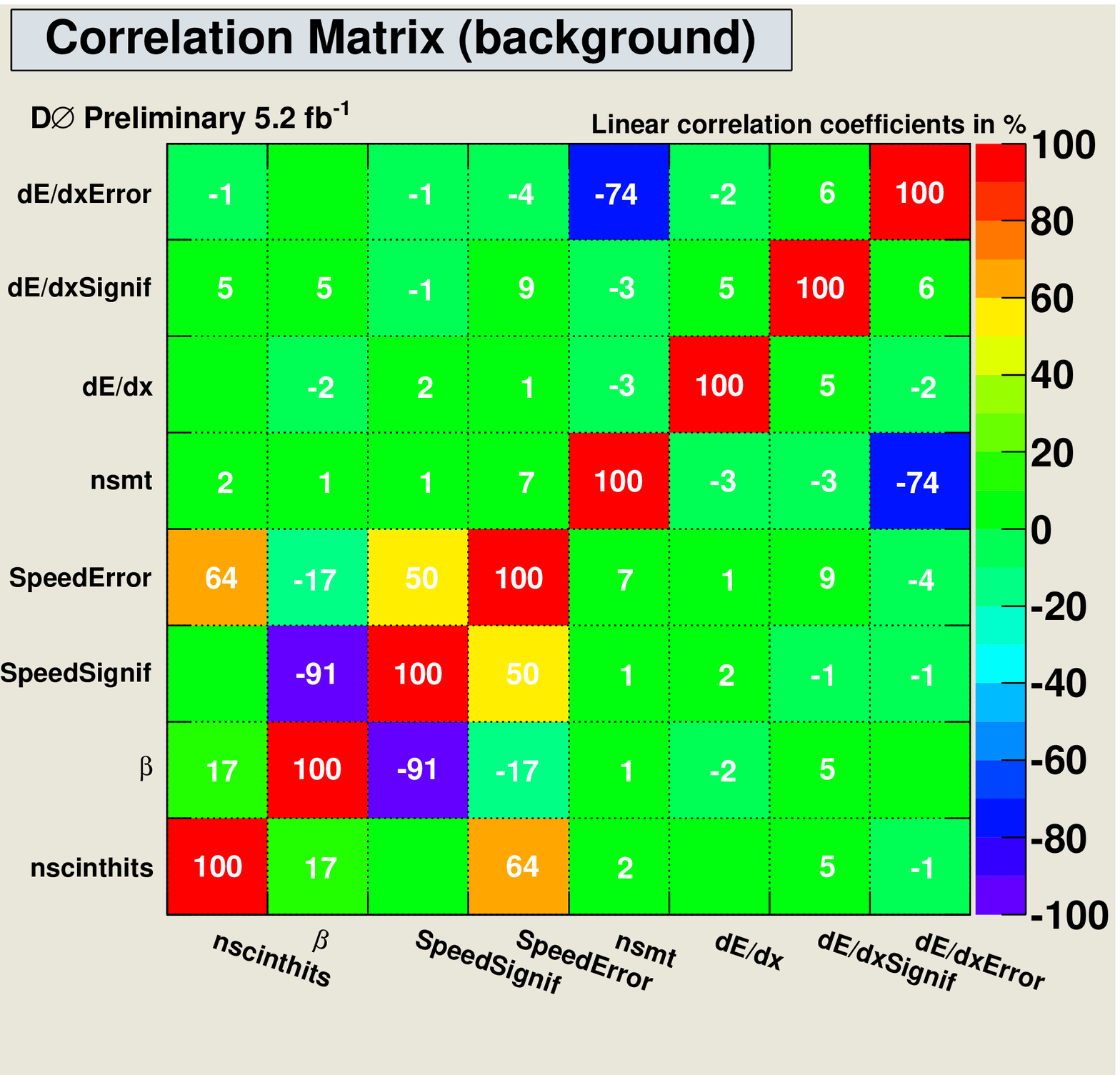}
\includegraphics[scale=0.4]{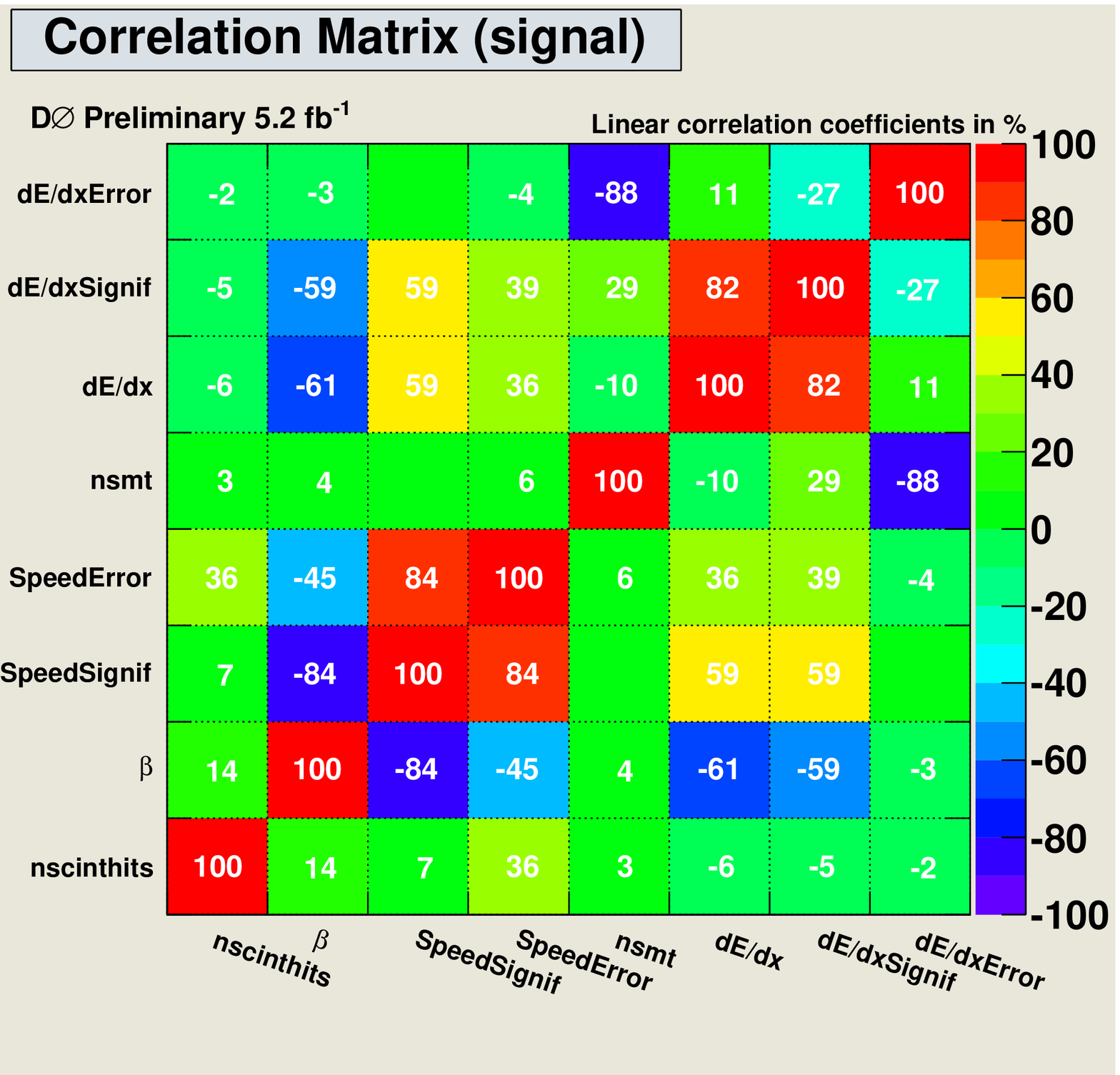}
\par\end{centering}
\caption{\label{fig:correlation_matrices} BDT correlation matrices
for background and a representative signal (300 GeV staus).}
\end{figure*}

\begin{figure*}[tbp]
\begin{centering}
\includegraphics[scale=0.4]{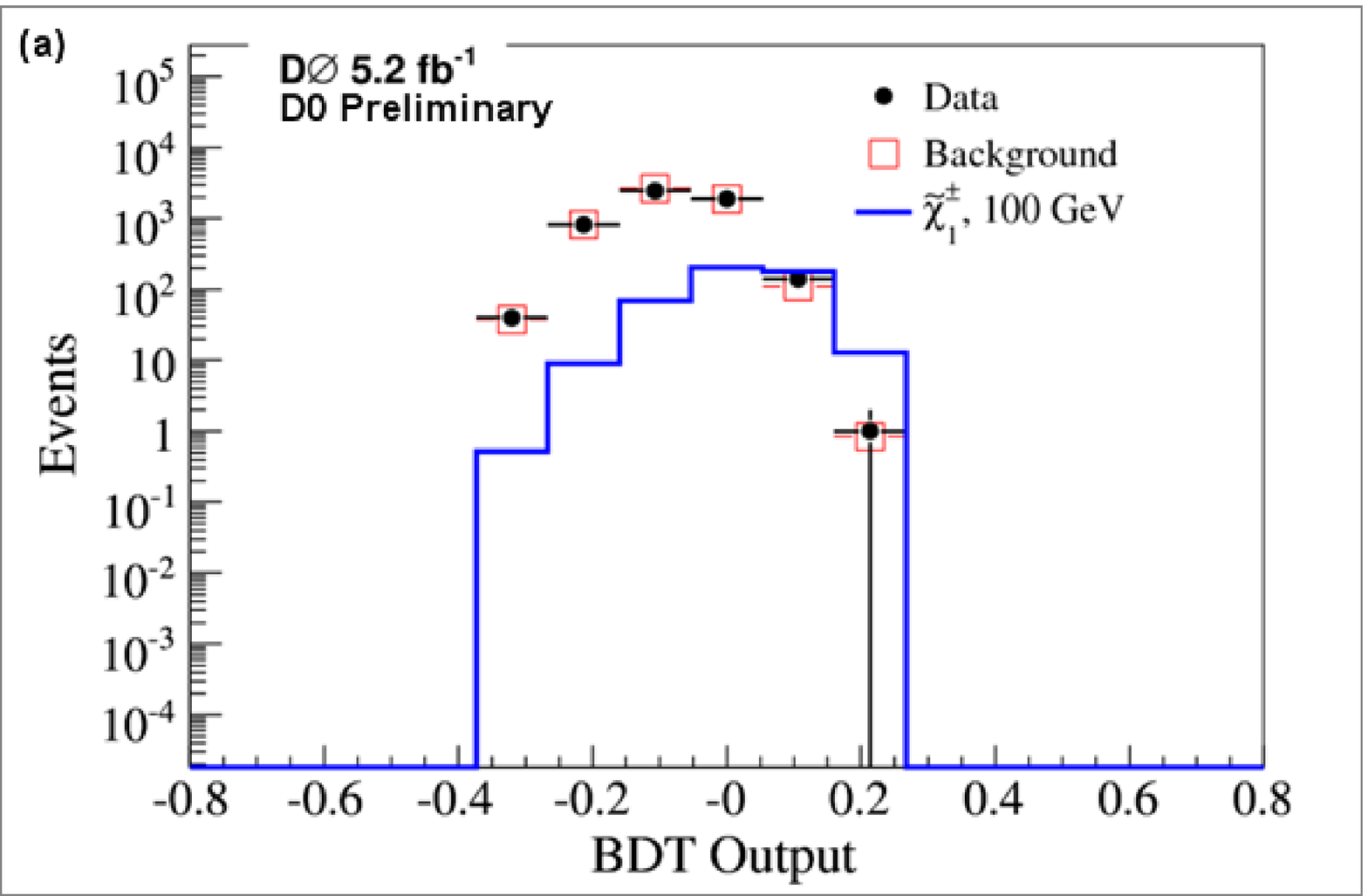}
\includegraphics[scale=0.4]{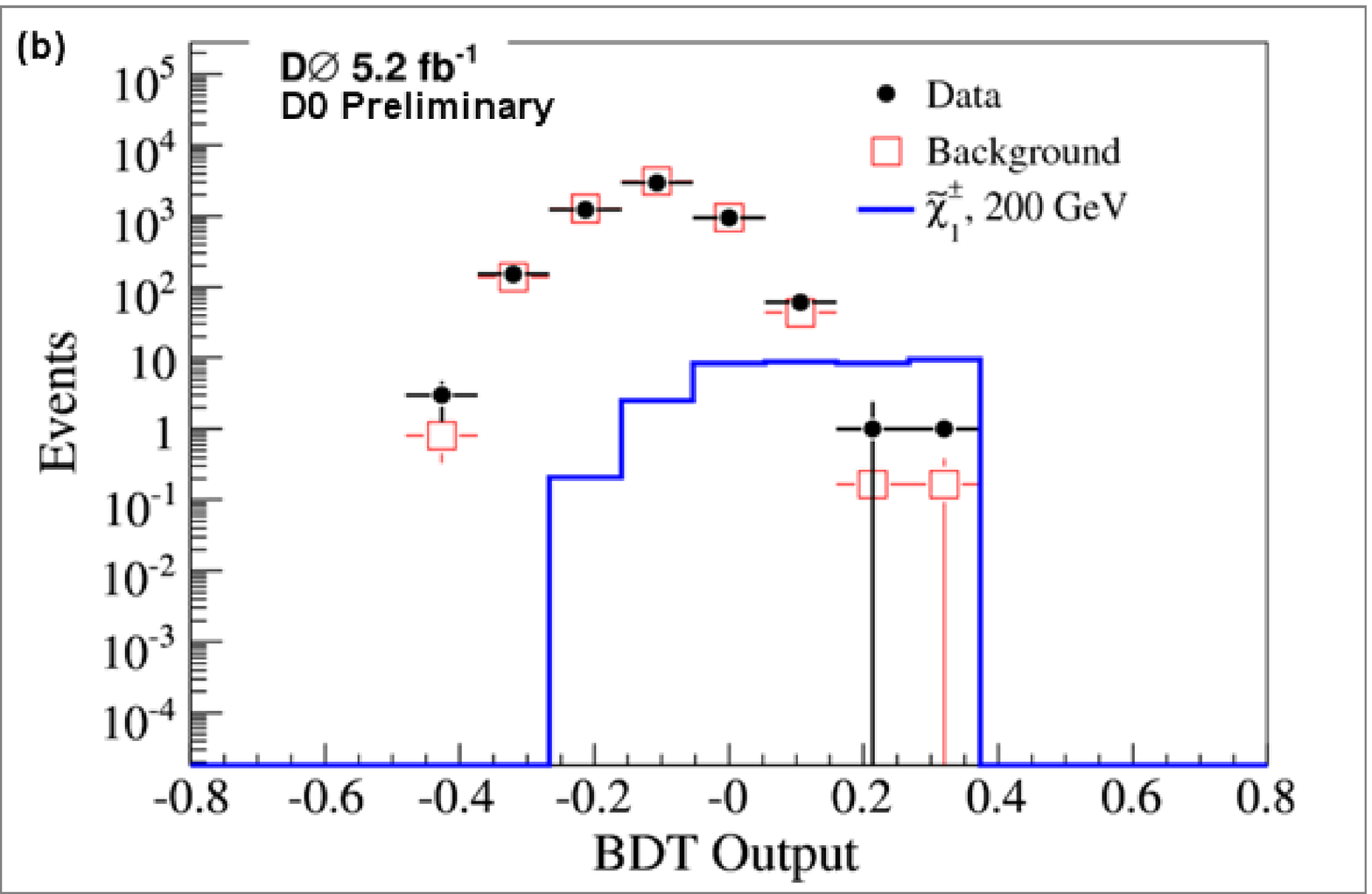}
\includegraphics[scale=0.4]{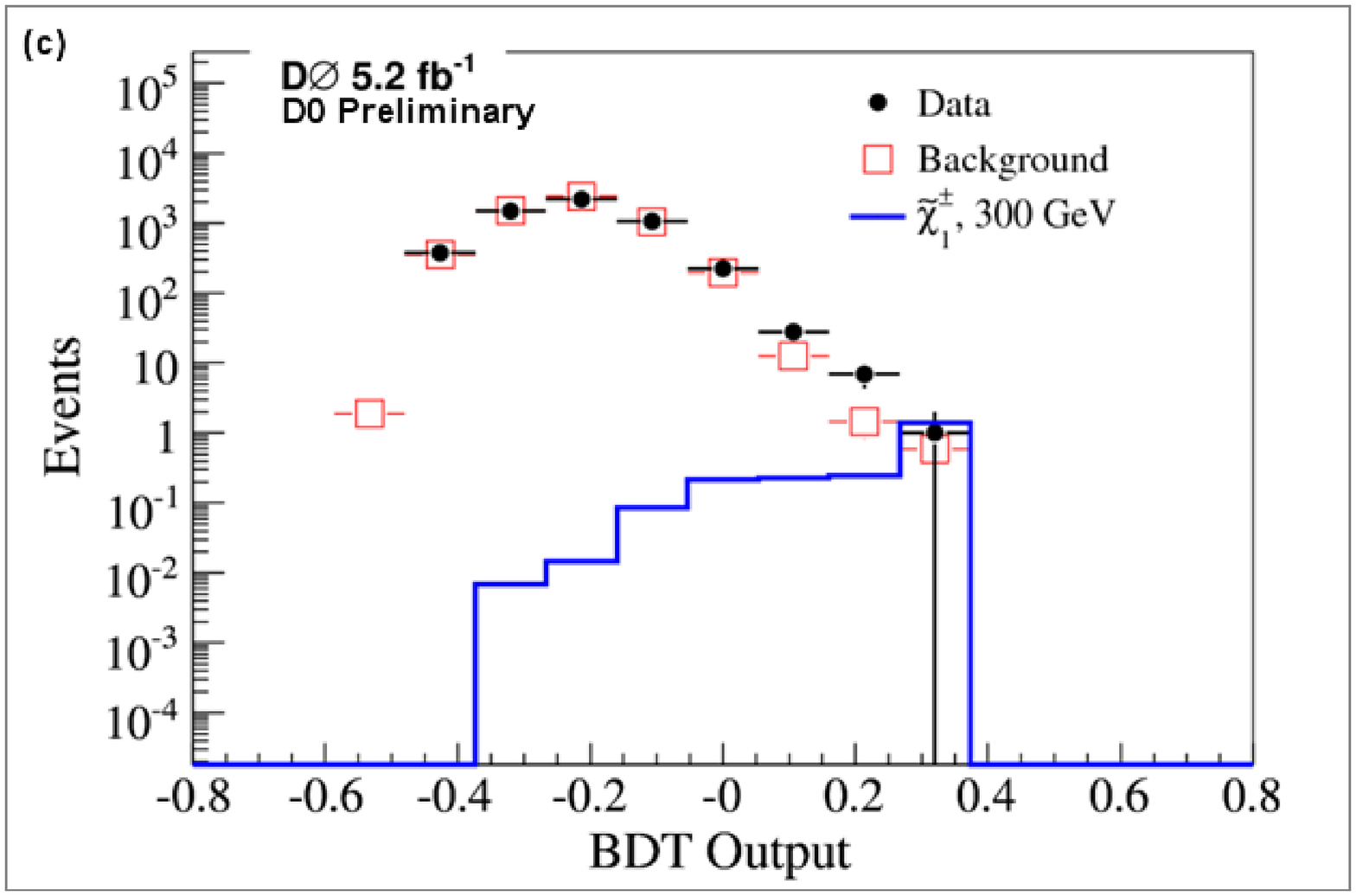}
\par\end{centering}
\caption{\label{fig:BDTdist_charginog} Final BDT distributions for signal,
background, and data. For the 100 GeV (a), 200 GeV (b), and 300 GeV (c) gaugino-like chargino cases. For each plot,
the signal histograms have been normalized to the expected number
of events.}
\end{figure*}

We utilize a boosted decision tree (BDT)~\cite{TMVA} to discriminate
signal from background. The discriminating variables are the CMLLP
candidate's $\beta$ and $dE/dx$, as well as several related variables:
 the speed significance, defined as $(1-\beta)/\sigma_{\beta}$,
the corresponding number of scintillator hits, the energy loss significance
defined as $(dE/dx-1)/\sigma_{dE/dx}$, and the number of SMT hits.
For each mass point in all four signal models we train the BDT with
the signal MC and the background, and then apply it to the data samples.
Figure~\ref{fig:beta_dedx} shows the distributions in $\beta$ and
in $dE/dx$ for the data and background samples, as well as for a
signal (gaugino-like charginos of mass 100 and 300 GeV). 
Figures~\ref{fig:speed_finaldist} and~\ref{fig:dedx_finaldist} show the 
discriminating variables that are input to the BDT, for data, background, and signal. 
Figure~\ref{fig:correlation_matrices} shows the correlations between input variables to the BDT, 
for background and a representative signal. 
Figure~\ref{fig:BDTdist_charginog} shows the BDT output distributions for three representative
signals: 100, 200, and 300 GeV gaugino-like charginos.

\section{Results}\label{sec:results}

Systematic uncertainties are studied by applying variations to the
data, background and signal samples and determining the deviations
in the BDT output distributions. Two systematic uncertainties affect
the shape of the BDT distributions, and their effect is taken into account explicitly in
the limit calculation: the uncertainty due to the width of the Level
1 trigger gate and the uncertainty of the time simulation in MC. Both of these uncertainties
are applied only to signal. By examining the signal-like region of
the BDT distributions, we find that the maximum (average) uncertainty
is 10\% (4\%) for the trigger gate width and 38\% (7\%) for the time
smearing. Other systematic uncertainties affect only the normalization of the BDT output.
The systematic uncertainties on the background are due to the $dE/dx$
correction ($< 0.1$\%) and the background normalization (7.5\%) The systematic
uncertainties on the signal include muon identification (2\%) and
the integrated luminosity (6.1\%)~\cite{d0lumi}. The systematic uncertainties
associated with the muon $p_{T}$$ $ smearing, the $dE/dx$ smearing
and the choice of PDF and factorization scale are all below 1\%.

We obtain the 95\% C.L. cross section limits from the BDT output distributions, constraining systematic uncertainties to data in background dominated
regions~\cite{Collie}.
These limits are shown in Tables~\ref{tab:limits_stau},~\ref{tab:limits_stop},~\ref{tab:limits_charginog}, and~\ref{tab:limits_charginoh}, together with
the NLO theoretical signal cross sections, computed with {\sc prospino}
~\cite{Prospino}. The limits vary from 0.04 pb to 0.006 pb for directly
pair-produced stau leptons with masses between 100 and 300 GeV. We
place similar bounds on the cross sections of other pair-produced
CMLLPs. Using the theoretical cross sections,
we are able to exclude gaugino-like charginos below 251 GeV
and higgsino-like charginos below 230 GeV. For stop quarks,
we assume a charge survival probability of 38\%, as discussed above,
and exclude masses below 265 GeV. If charge flipping does not
occur, we would obtain a significantly higher mass limit.

As seen in Tables~\ref{tab:limits_stau},~\ref{tab:limits_stop},~\ref{tab:limits_charginog}, and~\ref{tab:limits_charginoh},
 the observed limit exceeds the expected
limit at various mass points by as much as 2.5 standard deviations,
for all signals tested, due to the presence of the same
few data events with high BDT discriminant values. This discrepancy
reflects the excesses of data compared to background
observed in Fig.~\ref{fig:beta_dedx} for the distributions both in beta (around
0.6) and $dE/dx$ (around 2.8). We have compared these events
with background events and conclude that the observed excess
is consistent with a statistical fluctuation in the number
of such events.

\begin{table}[H]
\noindent \begin{centering}
\caption{\label{tab:limits_stau}NLO cross-sections and cross-section limits
for staus.}
\begin{tabular}{|c|c|c|c|}
\hline 
Mass [GeV] & NLO Cross-Section {[}pb{]} & 95\% CL Limit {[}pb{]} & Expected Limit \textbf{$\pm1\sigma$ }{[}pb{]}\tabularnewline
\hline
100 & 0.0121 & 0.0400 & 0.0263$_{-0.0075}^{+0.0109}$\tabularnewline
150 & 0.00214 & 0.0418 & 0.0164$_{-0.0035}^{+0.0062}$\tabularnewline
200 & 0.0004799 & 0.0113 & 0.00671$_{-0.00061}^{+0.00122}$\tabularnewline
250 & 0.000122 & 0.0132 & 0.00556$_{-0.00077}^{+0.00114}$\tabularnewline
300 & 0.0000314 & 0.00581 & 0.00538$_{-0.00076}^{+0.00104}$\tabularnewline
\hline
\end{tabular}
\par\end{centering}
\end{table}

\begin{table}[H]
\noindent \begin{centering}
\caption{\label{tab:limits_stop}NLO cross-sections and cross-section limits
for stops, assuming a charge survival probability of 38\%.}
\begin{tabular}{|c|c|c|c|}
\hline 
Mass [GeV] & NLO Cross-Section {[}pb{]} & 95\% CL Limit {[}pb{]} & Expected Limit \textbf{$\pm1\sigma$ }{[}pb{]}\tabularnewline
\hline
100 & 15.6 & 0.562 & 0.218$_{-0.062}^{+0.078}$\tabularnewline
150 & 1.58 & 0.113 & 0.0490$_{-0.0111}^{+0.0190}$\tabularnewline
200 & 0.266 & 0.0529 & 0.0234$_{-0.0037}^{+0.0106}$\tabularnewline
250 & 0.0560 & 0.0269 & 0.0201$_{-0.0090}^{+0.0037}$\tabularnewline
300 & 0.0130 & 0.0794 & 0.0529$_{-0.0128}^{+0.0140}$\tabularnewline
\hline
\end{tabular}
\par\end{centering}
\end{table}

\begin{table}[H]
\noindent \begin{centering}
\caption{\label{tab:limits_charginog}NLO cross-sections and cross-section
limits for gaugino-like charginos.}
\begin{tabular}{|c|c|c|c|}
\hline 
Mass [GeV] & NLO Cross-Section {[}pb{]} & 95\% CL Limit {[}pb{]} & Expected Limit \textbf{$\pm1\sigma$ }{[}pb{]}\tabularnewline
\hline
100 & 1.33 & 0.387 & 0.153$_{-0.043}^{+0.068}$\tabularnewline
150 & 0.235 & 0.0435 & 0.0167$_{-0.0033}^{+0.0054}$\tabularnewline
200 & 0.0566 & 0.0195 & 0.00945$_{-0.00057}^{+0.00368}$\tabularnewline
250 & 0.0153 & 0.0136 & 0.00988$_{-0.00127}^{+0.00402}$\tabularnewline
300 & 0.00417 & 0.0741 & 0.0185$_{-0.0027}^{+0.0046}$\tabularnewline
\hline
\end{tabular}
\par\end{centering}
\end{table}

\begin{table}[H]
\noindent \begin{centering}
\caption{\label{tab:limits_charginoh}NLO cross-sections and cross-section
limits for higgsino-like charginos.}
\begin{tabular}{|c|c|c|c|}
\hline 
Mass [GeV] & NLO Cross-Section {[}pb{]} & 95\% CL Limit {[}pb{]} & Expected Limit \textbf{$\pm1\sigma$ }{[}pb{]}\tabularnewline
\hline
100 & 0.381 & 0.106 & 0.110$_{-0.032}^{+0.050}$\tabularnewline
150 & 0.0736 & 0.0417 & 0.0165$_{-0.0038}^{+0.0053}$\tabularnewline
200 & 0.0186 & 0.0128 & 0.00852$_{-0.00112}^{+0.00169}$\tabularnewline
250 & 0.00525 & 0.00897 & 0.00716$_{-0.00100}^{+0.00267}$\tabularnewline
300 & 0.00154 & 0.0174 & 0.0119$_{-0.0005}^{+0.0033}$\tabularnewline
\hline
\end{tabular}
\par\end{centering}
\end{table}

\section{Summary}\label{sec:summary}

In summary, we have performed a search for charged, massive long-lived
particles using $5.2$ fb$^{\text{-1}}$ of integrated luminosity
with the D0 detector. We find no evidence of signal and set 95\% C.L.
cross-section upper limits which vary from 0.04 pb to 0.006 pb for
pair-produced stau leptons with masses in the range between 100 and
300 GeV. At 95\% C.L. we exclude pair-produced long-lived stop quarks
with mass below 265 GeV, gaugino-like charginos below 251 GeV, and
higgsino-like charginos below 230 GeV. These are presently the most
restrictive limits for chargino CMLLPs, with about a factor of five
improvement over the previous D0 cross section limits~\cite{D0 MSP PRL}.

\bigskip 
\begin{acknowledgments}
We thank the staffs at Fermilab and collaborating institutions, and
acknowledge support from the DOE and NSF (USA); CEA and CNRS/IN2P3
(France); FASI, Rosatom and RFBR (Russia); CNPq, FAPERJ, FAPESP and
FUNDUNESP (Brazil); DAE and DST (India); Colciencias (Colombia); CONACyT
(Mexico); KRF and KOSEF (Korea); CONICET and UBACyT (Argentina); FOM
(The Netherlands); STFC and the Royal Society (United Kingdom); MSMT
and GACR (Czech Republic); CRC Program and NSERC (Canada); BMBF and
DFG (Germany); SFI (Ireland); The Swedish Research Council (Sweden);
and CAS and CNSF (China).
\end{acknowledgments}

\bigskip 

\end{document}